\documentclass{aa}
\usepackage{times}
\usepackage{graphicx}
\begin{document}
   \title{Inelastic H+Li and H$^-$+Li$^+$ collisions and non-LTE Li I line formation in stellar atmospheres}

   \titlerunning{Inelastic H+Li and H$^-$+Li$^+$ collisions in stellar atmospheres}

   \author{P. S. Barklem
          \inst{1}
          \and
          A. K. Belyaev \inst{2}
          \and 
          M. Asplund \inst{3}
          }

   \offprints{\email{barklem@astro.uu.se}}

   \institute{Department of Astronomy and Space Physics, Uppsala University, Box
 515 SE 751 20 Uppsala, Sweden
         \and
             Department of Theoretical Physics, A. I. Herzen University, St.
 Petersburg 191186, Russia
         \and
             Research School of Astronomy, Mt.\ Stromlo Observatory, Cotter
 Road, Weston, ACT 2611, Australia
              }

   \date{Received ; accepted }

   \abstract{Rate coefficients for inelastic collisions between Li and H atoms covering all transitions between the asymptotic states Li($2s$,$2p$,$3s$,$3p$,$3d$,$4s$,$4p$,$4d$,$4f$)+H($1s$) and Li$^+$+H$^-$ are presented for the temperature range 2000--8000~K based on recent cross-section calculations. The data are of sufficient completeness for non-LTE modelling of the Li I 670.8\,nm and 610.4\,nm features in late-type stellar atmospheres.  Non-LTE radiative transfer calculations in both 1D and 3D model atmospheres have been carried out for test cases of particular interest. Our detailed calculations show that the classical modified Drawin-formula for collisional excitation and de-excitation (Li$^*$+H$\rightleftharpoons$Li$^{*\prime}$+H) over-estimates the cross-sections by typically several orders of magnitude and consequently that these reactions are negligible for the line formation process. However, the charge transfer reactions collisional ion-pair production and mutual neutralization (Li$^*$+H$\rightleftharpoons$Li$^+$+H$^-$) are of importance in thermalizing Li.  In particular, 3D non-LTE calculations of the Li I 670.8\,nm line in metal-poor halo stars suggest that 1D non-LTE results over-estimate the Li abundance by up to about 0.1 dex, aggrevating the discrepancy between the observed Li abundances and the primordial Li abundance as inferred by the WMAP analysis of the cosmic microwave background.
   }

   \maketitle
%

\section{Introduction}

Lithium abundances in stellar atmospheres are key observational parameters in astrophysics.  They tell us much about stellar evolution and nucleosynthesis, Big Bang nucleosynthesis, and the role of cosmic rays in Galactic chemical
 evolution, to name a few (see Lambert~\cite{lambert}; Carlsson~et~al.~\cite{carlsson} and references therein). Stellar Li abundances are in practice determined from the 670.8\,nm resonance line of Li~I and in a few cases the 610.4\,nm subordinate line.  Thus, the ability to accurately interpret these lines is important.  As Li~I has a low ionization potential, non-local thermodynamic equilibrium (non-LTE) effects can be expected to be important (Steenbock \&\ Holweger~\cite{steenbock_holweger}; Carlsson~et~al.~\cite{carlsson}; Asplund~et~al.~\cite{asplund}).

Steenbock and Holweger~(\cite{steenbock_holweger}) were the first to suggest that inelastic collisions with hydrogen might be important in thermalizing Li I.  In photospheres of late-type stars neutral hydrogen atoms outnumber electrons by typically $N_H/N_e\sim 10^4$, even more in metal poor stars.  Thus, weight of numbers may overcome the expectation that electron collisions are a more efficient thermalizing mechanism firstly due to their higher thermal velocity and therefore higher collision rate, and secondly since for speeds of interest electron cross sections are generally larger than those for neutrals since the neutral collisions are adiabatic while the electron collisions are not (e.g. Lambert~\cite{lambert}).  In such photospheres $kT\sim 0.2$--$0.6$~eV and thus for typical optical transitions the low energy collisions just above the threshold are most important in determining the collision rates.  Steenbock \&\ Holweger, in the absence of any better experimental or theoretical guidance, calculated the collisional excitation and ionization rates by modifying a formula from Drawin~(\cite{drawin}) for H+H which is itself a modification of the classical Thomson formula for excitation by electrons (cf.~Lambert~\cite{lambert}). In recent years some work has been devoted to improve these highly uncertain H collision cross-sections. Low-energy experimental data have been obtained (Fleck et~al.~\cite{fleck}) and {\it ab initio} calculations performed down to the threshold (Belyaev~et~al.~\cite{belyaev99}) for the Na+H system with good agreement between theory and experiment.  However, these results are for only the very lowest states of this one system.  For non-LTE calculations, estimates are required for transitions between all states which might affect the population of the states of interest.  It is for this reason that the modified Drawin formula (often with a scaling factor determined astrophysically) is still often in use despite that it is known to overestimate the Na(3s)+H$\rightarrow$Na(3p)+H collision rate by several orders of magnitude.

Recently, calculations for Li+H collisions for all transitions between the states Li($2s$,$2p$,$3s$,$3p$,$3d$,$4s$,$4p$,$4d$,$4f$)+H($1s$) and Li$^+$+H$^-$ have been performed (Belyaev \&\ Barklem~\cite{paper1}).  This paper should be consulted for details and justification of the approaches used.  In brief summary, on the basis of the interpretation of the experimental and theoretical results for Na+H (Fleck~et~al.~\cite{fleck}; Belyaev~et al.~\cite{belyaev99}), non-adiabatic regions associated with avoided ionic crossings in the adiabatic potential curves are expected to provide the basic mechanism for the transitions at low energy.  A simple physical interpretation of this process is that at certain separations during the collision the optical electron associated with the Li atom may tunnel to the H atom, leading to a predominantly ionic charge distribution (Li$^+$H$^-$) of the LiH quasi-molecule.  Later, at another avoided crossing, the electron may tunnel back towards the Li atom into a different covalent charge distribution leading asymptotically to a different final Li state. The electron may also remain with the H atom leading to an ion-pair production reaction (Li$^*$+H$\rightarrow$Li$^+$+H$^-$).  For the transitions between the lowest four states a fully quantum mechanical approach has been used, while for the transitions involving higher states, a multichannel Landau-Zener model has been used.  The model approach is justified for the higher crossings where the requirements of the Landau-Zener model are better fulfilled.

For the charge transfer reactions Li$^*$+H$\rightleftharpoons$Li$^+$+H$^-$, referred to as ion-pair production and mutual neutralization, our calculations use the model approach.  More elaborate calculations, using a diabatic molecular expansion method, have been made by Croft et~al.~(\cite{croft_mnras}).  This process has a remarkably large cross section near the threshold in alkali-hydrogen systems, particularly to and from the first excited $s$ state.  This has been predicted theoretically (Bates \&\ Boyd~\cite{bates_boyd}; Janev \&\ Radulovi\'c~\cite{janev}; Croft et~al.~\cite{croft_jphysb}; Dickinson~et~al.~\cite{dickinson}) and is consistent with experiment (Peart \&\ Hayton~\cite{peart_hayton}). Despite that this process has been known to be remarkably efficient in various fields (e.g. plasma physics, early universe chemistry) for some time, to our knowledge this is the first time it has been considered in a stellar atmosphere application of Li.

\section{Collision rate coefficients}

The rate coefficients $\langle \sigma v \rangle$ for excitation and de-excitation H($1s$)+Li($nl$)$\rightarrow$H($1s$)+Li($n^\prime l^\prime$), and the charge transfer reactions Li$(nl)$+H$\rightleftharpoons$Li$^+$+H$^-$ are presented in Table~\ref{tab:rates}.  The data have been obtained by integrating a Maxwellian velocity distribution with cross-sections from Belyaev \&\ Barklem~(\cite{paper1}) in the case of excitation and de-excitation, and from Croft et~al.~(\cite{croft_mnras}) in the case of mutual neutralization, with ion-pair production cross sections obtained via the principle of detailed balance.  Thus the data represent, in our opinion, the best available estimates. Note that all forward and backward rates were computed separately, but should be connected by the detailed balance relation. Inspection of the data reveals the remarkably large rate coefficients for Li$(nl)$+H$\rightleftharpoons$Li$^+$+H$^-$, particularly for the $n=3$ states.

\begin{table*}
\caption{Rate coefficients $\langle \sigma v \rangle$, in units of cm$^3$~s$^{-1}$, for the processes H($1s$)+Li($nl$)$\rightarrow$H($1s$)+Li($n^\prime l^\prime$), or where indicated H($1s$)+Li($nl$)$\rightarrow$H$^-$+Li$^+$($1s^2$) and H$^-$+Li$^+$($1s^2$)$\rightarrow$H($1s$)+Li($n^\prime l^\prime$), for selected temperatures in the range $T=2000$--8000 K. Asterisks indicate transitions where the calculated cross sections are considered unreliable.} 
\label{tab:rates}
\tiny
\begin{center}
\begin{tabular}{ccccccccccc}
\hline
Initial & \multicolumn{9}{c}{final state $n^\prime l^\prime$}  \\
state $nl$   & 2s & 2p & 3s & 3p & 3d & 4s & 4p & 4d & 4f & H$^-$+Li$^+$  \\
\hline
\multicolumn{11}{c}{\underline{2000K}} \\
          2s & ---          & $ 9.18$E$-20$  & $ 1.20$E$-25$  & $ 1.67$E$-27$  & $ 4.16$E$-28$  & $ 2.18$E$-29$  & $ 2.19$E$-30$  & $ 1.85$E$-30$  & $ 1.84$E$-30$  & $ 2.11$E$-25$ \\
          2p & $ 1.39$E$-15$  & ---          & $ 8.31$E$-18$  & $ 2.76$E$-19$  & $ 7.91$E$-20$  & $ 6.23$E$-21$  & $ 7.04$E$-22$  & $ 4.66$E$-22$  & $ 3.26$E$-22$  & $ 2.06$E$-19$ \\
          3s & $ 3.79$E$-17$  & $ 1.74$E$-13$  & ---          & $ 2.02$E$-11$  & $ 3.77$E$-12$  & $ 1.88$E$-12$  & $ 1.78$E$-13$  & $ 1.15$E$-13$  & $ 2.51$E$-14$  & $ 1.75$E$-11$ \\
          3p & $ 2.56$E$-18$  & $ 2.80$E$-14$  & $ 9.78$E$-11$  & ---          & $ 2.23$E$-10$  & $ 3.55$E$-11$  & $ 3.30$E$-12$  & $ 2.21$E$-12$  & $ 2.27$E$-13$  & $ 4.25$E$-11$ \\
          3d & $ 4.91$E$-19$  & $ 6.22$E$-15$  & $ 1.42$E$-11$  & $ 1.72$E$-10$  & ---          & $ 2.92$E$-12$  & $ 2.74$E$-13$  & $ 1.83$E$-13$  & $ 1.63$E$-14$  & $ 3.27$E$-12$ \\
          4s & $ 1.70$E$-18$  & $ 3.53$E$-14$  & $ 5.15$E$-10$  & $ 1.97$E$-09$  & $ 2.10$E$-10$  & ---          &    *           &  *             &  *             & $ 3.42$E$-13$ \\
          4p & $ 1.78$E$-19$  & $ 2.97$E$-15$  & $ 4.51$E$-11$  & $ 1.58$E$-10$  & $ 1.72$E$-11$  &  *             & ---          &  *           &  *           & $ 3.95$E$-14$ \\
          4d & $ 9.07$E$-20$  & $ 1.30$E$-15$  & $ 1.89$E$-11$  & $ 6.67$E$-11$  & $ 7.31$E$-12$  &  *             &  *           & ---          &  *           &  *          \\
          4f & $ 2.69$E$-20$  & $ 5.87$E$-16$  & $ 3.10$E$-12$  & $ 5.49$E$-12$  & $ 5.10$E$-13$  &  *             &  *           &  *           & ---          &  *          \\
H$^-$+Li$^+$ & $ 4.12$E$-13$  & $ 2.65$E$-11$  & $ 1.07$E$-07$  & $ 5.39$E$-08$  & $ 5.34$E$-09$  & $ 7.63$E$-12$  & $ 9.30$E$-13$  &  *           &  *           & ---         \\
\multicolumn{11}{c}{\underline{4000K}} \\
          2s & ---          & $ 3.62$E$-17$  & $ 2.70$E$-21$  & $ 9.94$E$-23$  & $ 1.58$E$-23$  & $ 2.30$E$-24$  & $ 3.88$E$-25$  & $ 3.46$E$-25$  & $ 3.45$E$-25$  & $ 4.62$E$-19$ \\
          2p & $ 2.57$E$-15$  & ---          & $ 1.23$E$-15$  & $ 7.06$E$-17$  & $ 1.34$E$-17$  & $ 3.10$E$-18$  & $ 5.86$E$-19$  & $ 4.09$E$-19$  & $ 2.87$E$-19$  & $ 7.92$E$-16$ \\
          3s & $ 4.79$E$-17$  & $ 3.09$E$-13$  & ---          & $ 5.95$E$-11$  & $ 9.15$E$-12$  & $ 1.16$E$-11$  & $ 1.79$E$-12$  & $ 1.21$E$-12$  & $ 2.66$E$-13$  & $ 5.53$E$-10$ \\
          3p & $ 2.25$E$-18$  & $ 2.25$E$-14$  & $ 7.56$E$-11$  & ---          & $ 1.26$E$-10$  & $ 5.96$E$-11$  & $ 8.73$E$-12$  & $ 6.14$E$-12$  & $ 6.36$E$-13$  & $ 3.42$E$-10$ \\
          3d & $ 2.42$E$-19$  & $ 2.91$E$-15$  & $ 7.93$E$-12$  & $ 8.56$E$-11$  & ---          & $ 4.35$E$-12$  & $ 6.38$E$-13$  & $ 4.49$E$-13$  & $ 4.02$E$-14$  & $ 2.39$E$-11$ \\
          4s & $ 6.07$E$-19$  & $ 1.27$E$-14$  & $ 1.92$E$-10$  & $ 7.62$E$-10$  & $ 8.16$E$-11$  & ---          &   *            &  *             &  *             & $ 1.47$E$-12$ \\
          4p & $ 6.33$E$-20$  & $ 1.06$E$-15$  & $ 1.61$E$-11$  & $ 5.69$E$-11$  & $ 6.19$E$-12$  &  *             & ---          &  *           &  *           & $ 1.54$E$-13$ \\
          4d & $ 3.23$E$-20$  & $ 4.63$E$-16$  & $ 6.75$E$-12$  & $ 2.39$E$-11$  & $ 2.62$E$-12$  &  *             &  *           & ---          &  *           &  *          \\
          4f & $ 9.60$E$-21$  & $ 2.09$E$-16$  & $ 1.11$E$-12$  & $ 1.98$E$-12$  & $ 1.84$E$-13$  &  *             &  *           &  *           & ---          &  *          \\
H$^-$+Li$^+$ & $ 1.29$E$-12$  & $ 3.11$E$-11$  & $ 8.67$E$-08$  & $ 4.22$E$-08$  & $ 4.33$E$-09$  & $ 1.39$E$-11$  & $ 2.60$E$-12$  &  *           &  *           & ---         \\
\multicolumn{11}{c}{\underline{6000K}} \\
          2s & ---          & $ 4.18$E$-16$  & $ 1.17$E$-19$  & $ 5.33$E$-21$  & $ 5.90$E$-22$  & $ 8.37$E$-23$  & $ 1.68$E$-23$  & $ 1.52$E$-23$  & $ 1.52$E$-23$  & $ 8.08$E$-17$ \\
          2p & $ 4.97$E$-15$  & ---          & $ 9.51$E$-15$  & $ 5.53$E$-16$  & $ 7.32$E$-17$  & $ 1.89$E$-17$  & $ 4.24$E$-18$  & $ 3.02$E$-18$  & $ 2.12$E$-18$  & $ 1.55$E$-14$ \\
          3s & $ 7.96$E$-17$  & $ 5.45$E$-13$  & ---          & $ 8.30$E$-11$  & $ 1.08$E$-11$  & $ 1.64$E$-11$  & $ 2.96$E$-12$  & $ 2.04$E$-12$  & $ 4.49$E$-13$  & $ 1.72$E$-09$ \\
          3p & $ 2.96$E$-18$  & $ 2.58$E$-14$  & $ 6.75$E$-11$  & ---          & $ 8.37$E$-11$  & $ 5.46$E$-11$  & $ 9.29$E$-12$  & $ 6.64$E$-12$  & $ 6.91$E$-13$  & $ 6.59$E$-10$ \\
          3d & $ 2.13$E$-19$  & $ 2.23$E$-15$  & $ 5.74$E$-12$  & $ 5.44$E$-11$  & ---          & $ 3.82$E$-12$  & $ 6.51$E$-13$  & $ 4.65$E$-13$  & $ 4.18$E$-14$  & $ 4.67$E$-11$ \\
          4s & $ 3.32$E$-19$  & $ 6.94$E$-15$  & $ 1.06$E$-10$  & $ 4.27$E$-10$  & $ 4.59$E$-11$  & ---          &    *           &  *             &  *             & $ 3.22$E$-12$ \\
          4p & $ 3.46$E$-20$  & $ 5.76$E$-16$  & $ 8.81$E$-12$  & $ 3.12$E$-11$  & $ 3.39$E$-12$  & *            & ---          &  *           &  *           & $ 3.36$E$-13$ \\
          4d & $ 1.76$E$-20$  & $ 2.53$E$-16$  & $ 3.68$E$-12$  & $ 1.31$E$-11$  & $ 1.44$E$-12$  & *            &  *           & ---          &  *           &  *          \\
          4f & $ 5.24$E$-21$  & $ 1.14$E$-16$  & $ 6.04$E$-13$  & $ 1.08$E$-12$  & $ 1.01$E$-13$  & *            &  *           &  *           & ---          &  *          \\
H$^-$+Li$^+$ & $ 2.54$E$-12$  & $ 4.11$E$-11$  & $ 7.95$E$-08$  & $ 3.74$E$-08$  & $ 4.05$E$-09$  & $ 2.29$E$-11$  & $ 5.05$E$-12$  &  *           &  *           & ---         \\
\multicolumn{11}{c}{\underline{8000K}} \\
          2s & ---          & $ 1.90$E$-15$  & $ 1.11$E$-18$  & $ 5.49$E$-20$  & $ 4.86$E$-21$  & $ 4.44$E$-22$  & $ 9.73$E$-23$  & $ 8.91$E$-23$  & $ 8.89$E$-23$  & $ 1.16$E$-15$ \\
          2p & $ 9.27$E$-15$  & ---          & $ 3.24$E$-14$  & $ 1.84$E$-15$  & $ 1.92$E$-16$  & $ 4.12$E$-17$  & $ 1.00$E$-17$  & $ 7.22$E$-18$  & $ 5.06$E$-18$  & $ 7.81$E$-14$ \\
          3s & $ 1.48$E$-16$  & $ 8.89$E$-13$  & ---          & $ 9.90$E$-11$  & $ 1.13$E$-11$  & $ 1.71$E$-11$  & $ 3.36$E$-12$  & $ 2.34$E$-12$  & $ 5.14$E$-13$  & $ 3.04$E$-09$ \\
          3p & $ 4.77$E$-18$  & $ 3.28$E$-14$  & $ 6.44$E$-11$  & ---          & $ 6.14$E$-11$  & $ 4.60$E$-11$  & $ 8.44$E$-12$  & $ 6.08$E$-12$  & $ 6.34$E$-13$  & $ 9.02$E$-10$ \\
          3d & $ 2.69$E$-19$  & $ 2.19$E$-15$  & $ 4.73$E$-12$  & $ 3.91$E$-11$  & ---          & $ 3.16$E$-12$  & $ 5.79$E$-13$  & $ 4.17$E$-13$  & $ 3.76$E$-14$  & $ 6.62$E$-11$ \\
          4s & $ 2.16$E$-19$  & $ 4.52$E$-15$  & $ 6.97$E$-11$  & $ 2.82$E$-10$  & $ 3.03$E$-11$  & ---          &  *             &  *             &  *             & $ 5.37$E$-12$ \\
          4p & $ 2.25$E$-20$  & $ 3.75$E$-16$  & $ 5.73$E$-12$  & $ 2.03$E$-11$  & $ 2.21$E$-12$  &  *           & ---          &  *           &  *           & $ 5.59$E$-13$ \\
          4d & $ 1.15$E$-20$  & $ 1.64$E$-16$  & $ 2.40$E$-12$  & $ 8.51$E$-12$  & $ 9.35$E$-13$  &  *           &  *           & ---          &  *           &  *          \\
          4f & $ 3.41$E$-21$  & $ 7.40$E$-17$  & $ 3.93$E$-13$  & $ 7.06$E$-13$  & $ 6.59$E$-14$  &  *           &  *           &  *           & ---          &  *          \\
H$^-$+Li$^+$ & $ 3.89$E$-12$  & $ 5.37$E$-11$  & $ 7.61$E$-08$  & $ 3.47$E$-08$  & $ 3.98$E$-09$  & $ 3.30$E$-11$  & $ 7.94$E$-12$  &  *           &  *           & ---         \\
\hline
\end{tabular}
\end{center}
\end{table*}

For the temperatures here the uncertainty in the rate coefficients is determined by the uncertainty in the cross sections near the threshold. The uncertainty in the input cross sections is discussed in detail in the relevant papers. In summary, the accuracy for transitions between the lower states ($2s$,$2p$ and $3s$) is estimated to be better than a factor of 2. For the excitations involving higher states the data are expected to be accurate within an order of magnitude.  Comparison with experimental results for the total neutralization cross-section with D$^-$ (which uses the same interaction potentials as for H$^-$) finds the theoretical result about 20\% greater at the lowest measured energy 0.68 eV (see Croft et~al.~\cite{croft_jphysb}), with a general trend of the disagreement becoming larger at lower energies, recalling that the collision energies of importance here are $E\approx 0.2-0.6$~eV.

Lambert~(\cite{lambert}) presented unpublished estimated rate coefficients for the $2s$$\rightarrow$$2p$ transition computed by Allen \&\ Dickinson.  Our data are roughly an order of magnitude smaller.

Comparisons were made with the modified Drawin formula for all optically allowed transitions at 5000 K.  Except for an extreme outlier ($3s$$\rightarrow$$4p$), the Drawin formula gives a rate coefficient between roughly one and six orders of magnitude greater than our result.  This reinforces the view that the Drawin formula typically greatly over-estimates the collisional efficiency, as judged by the available (admittedly rather scant) evidence.

\section{Application to non-LTE formation of Li lines}

\begin{figure}
\begin{center}
\resizebox{75mm}{!}{\rotatebox{0}{\includegraphics{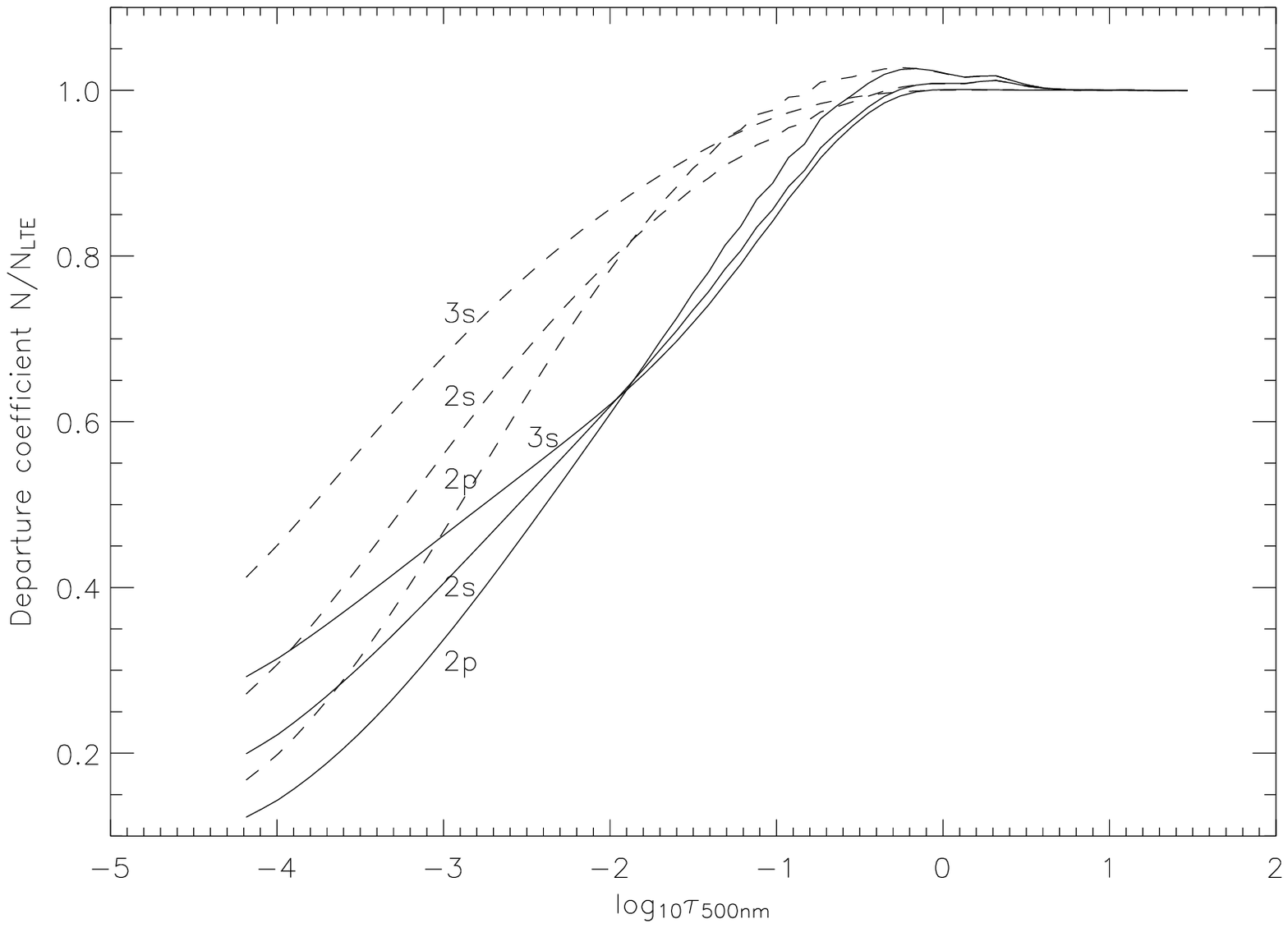}}}
\resizebox{75mm}{!}{\rotatebox{0}{\includegraphics{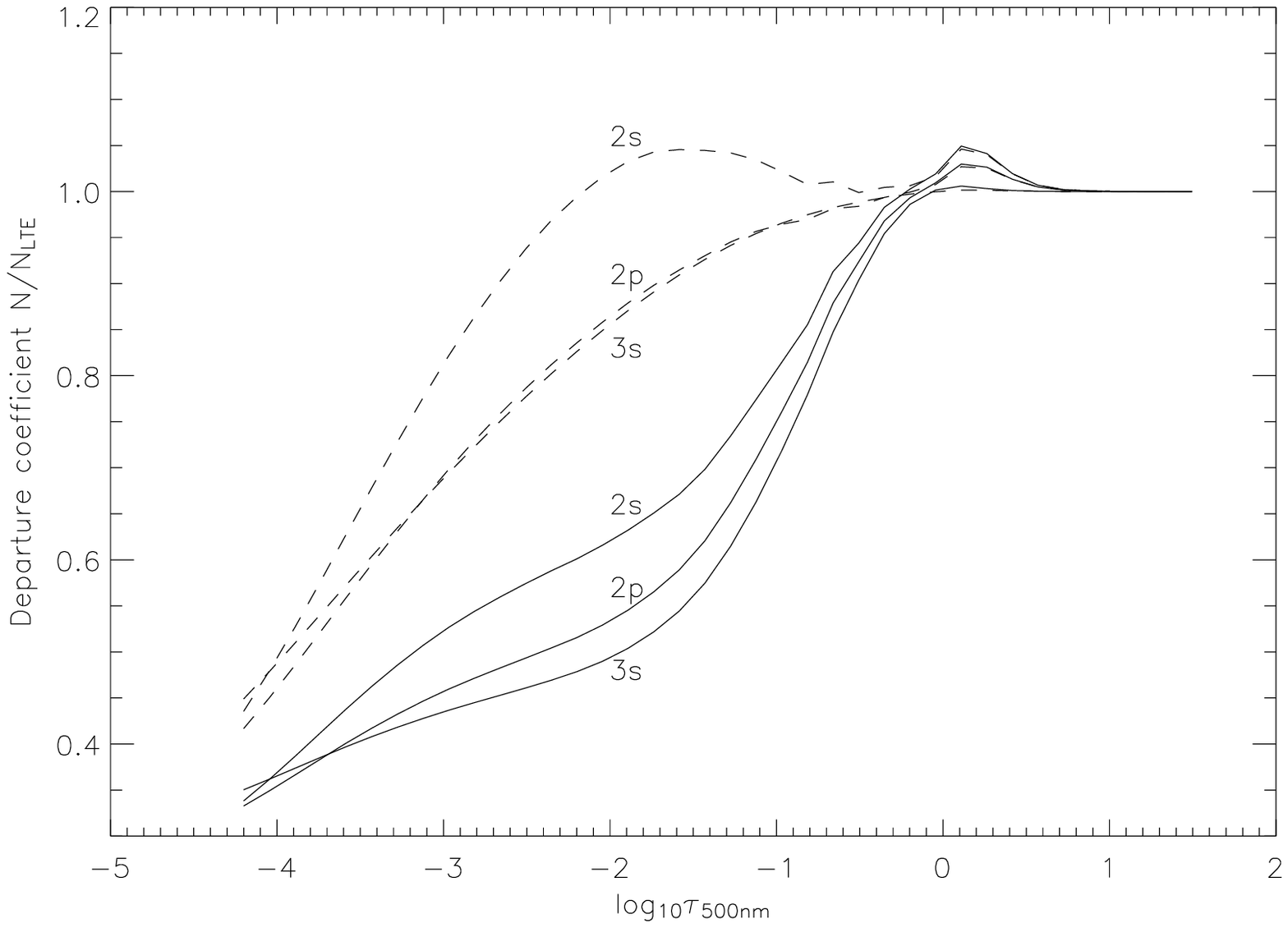}}}
\end{center}
\caption{Departure coefficients with standard optical depth for lower Li~I levels  in the solar 1D model with $\log \epsilon_\mathrm{Li}=1.1$ (upper panel) and in the HD140283 1D model with $\log \epsilon_\mathrm{Li}=1.8$ (lower panel). Full lines are for the atomic model without H collisions \texttt{noH} and the dashed lines are with H collisions \texttt{wH}.
}
\label{fig:1}
\end{figure}

To assess the impact of the data on the interpretation of stellar spectra, we have performed non-LTE calculations for Li in cases of particular interest using the \texttt{MULTI}-code for 1D statistical equilibrium problems (Carlsson~\cite{multi}) and the \texttt{MULTI3D}-code (Botnen~\cite{botnen}; Botnen \&\ Carlsson~\cite{botnen_carlsson}; Asplund~et~al.~\cite{asplund}) for the 3D case.  The adopted model atom is the 21-level atom compiled by Carlsson et~al.~(\cite{carlsson}). To test the impact of the new data three model atoms were used, namely the original model atom with no H collisions (\texttt{noH}), the model with the new H collision data added (\texttt{wH}), and a model with the H collision rates halved (\texttt{w0.5H}); the LTE results have been obtained with the same codes using an atom with extremely large collisional cross-sections to ensure consistency with the non-LTE calculations.  The 1D model atmospheres are from the \texttt{MARCS} code (Gustafsson et al.~\cite{marcs} and subsequent updates). The 3D models and calculations are as detailed in Asplund et al.~(\cite{asplund99}, \cite{asplund}). Calculations were performed for the Sun, the metal-poor subgiant HD140283 and the metal-poor turnoff star HD84937.  Equivalent widths for the 670.8\,nm line are presented in Table~\ref{tab:ews}. Departure coefficients for the lower states in the 1D non-LTE models for the Sun and HD140283 are presented in Fig.~\ref{fig:1}.

\begin{table*}
\caption{Predicted flux equivalent widths (in pm=10\,m\AA ) for the 670.8\,nm line in the Sun, HD140283 and HD84937 based on 1D and 3D models, in both LTE and non-LTE.  Results are shown for various model atoms as described in the text. Note that the 3D results shown are based on only one snapshot and hence are not appropriate for a direct comparison with the 1D cases.  }
\label{tab:ews}
\scriptsize
\begin{center}
\begin{tabular}{ccccccccc|cccc}
\hline
 & & & & & \multicolumn{4}{c|}{1D} &  \multicolumn{4}{c}{3D} \\
Star & \multicolumn{3}{c}{Parameters} & $\log \epsilon_\mathrm{Li}$ &
 $W_\lambda^\mathrm{LTE}$  & 
       $W_\lambda^\mathrm{NLTE}$  & $W_\lambda^\mathrm{NLTE}$  & 
       $W_\lambda^\mathrm{NLTE}$ &
       $W_\lambda^\mathrm{LTE}$  &
       $W_\lambda^\mathrm{NLTE}$  & $W_\lambda^\mathrm{NLTE}$  & 
       $W_\lambda^\mathrm{NLTE}$ \\
 & $T_\mathrm{eff}$ & $\log g$ & [Fe/H] & &           & \texttt{noH} &
 \texttt{wH} & \texttt{w0.5H} & & \texttt{noH} & \texttt{wH} & \texttt{w0.5H} \\
 & [K] & [cm s$^{-2}$] & & & [pm]          & [pm] & [pm]& [pm]& [pm] & [pm] & [pm] & [pm]\\
\hline
Sun      & 5777&4.44&0.0     & 1.1 & 0.40 & 0.34 & 0.38 & 0.37 & 0.55 & 0.37 & 0.40 &  0.39    \\
HD140283 & 5690&3.87&$-$2.5  & 1.8 & 2.40 & 2.18 & 2.66 & 2.61 & 3.84 & 1.96 & 2.35 & 2.32 \\
HD84937  & 6330&4.04&$-$2.25 & 2.0 & 1.31 & 1.44 & 1.57 & 1.55 & 1.79 & 1.11 & 1.26 &  1.24    \\
\hline
\end{tabular}
\end{center}
\end{table*}

Although quantitatively different, the overall effects of the H collisions are similar in the 1D and 3D cases. Almost all of the difference resulting from the inclusion of the H collision data is caused by the reaction Li$(3s)$+H$\rightleftharpoons$Li$^+$+H$^-$. Removing all H collisions but this reaction results in equivalent widths within 1\% of the results for the \texttt{wH} model.  This reaction causes a stronger collisional coupling between the ionization stages, nearly thermalizing the $3s$ state.  Photon suction (Bruls et al.~\cite{bruls}) results in the other levels being affected, in particular the ground state is pushed towards LTE, or even into overpopulation as seen in the case of HD140283 in 1D, at around $\log \tau_{500}=-2$.  As clear from Table~\ref{tab:ews}, the results are not particularly sensitive to the details of the H collisional cross-sections with only small differences resulting when decreasing all H cross-sections by 50\%\ (\texttt{w0.5H}). The remaining uncertainties in the charge transfer reactions are thus expected to have limited impact on the Li I line formation.

In terms of abundances, the inclusion of charge transfer reactions lowers the derived values by about 0.05--0.10\,dex for all three stellar models, whether in 1D or 3D.  While the 1D non-LTE abundance corrections (1D non-LTE -- 1D LTE) are small ($<$0.05\,dex) for the relevant stars (Table~\ref{tab:ews}, see also Carlsson~et~al.~\cite{carlsson}), the 3D non-LTE effects (3D non-LTE -- 3D LTE) are substantial for metal-poor stars (Asplund~et~al.~\cite{asplund}). However, since the differences between 3D LTE and 1D LTE are also significant and of opposite sign (Asplund~et~al.~\cite{asplund99}), the net difference between 3D non-LTE and 1D non-LTE is less dramatic. For the two metal-poor stars this difference amounts to about 0.05--0.10\,dex {\em lower} abundance than in 1D\footnote{Note that it is not appropriate to directly compare the equivalent widths in Table~\ref{tab:ews} to estimate the net abundance difference between 1D and 3D as the 3D results are here only shown for one snapshot. The proper procedure is to add the 3D non-LTE corrections obtained here to the 3D LTE corrections published by Asplund~et~al.~(\cite{asplund99}) based on a long time-sequence with better numerical resolution.}. This has a direct bearing on the question of the primordial Li abundance. The observational evidence suggests a plateau at $\log \epsilon_\mathrm{Li} = 2.1-2.3$ with 1D non-LTE abundance corrections taken into account (Ryan et al.~\cite{ryan}; Bonifacio~\cite{bonifacio}), with the abundance range partly reflecting different effective temperature scales for halo stars. By slightly decreasing these abundances, the inclusion of 3D models and charge transfer reactions slightly aggravates the discrepancy between the observed Li abundances and the predicted primordial Li abundance from Big Bang nucleosynthesis. High-redshift deuterium measurements and analyses of the cosmic microwave background yield a relatively high baryon density of the Universe (e.g. Burles et al.~\cite{burles}). In particular, the recent WMAP analysis (Bennett et al.~\cite{wmap}) corresponds to a primordial Li abundance of $\log \epsilon_\mathrm{Li,p} = 2.69\pm0.13$ based on standard Big Bang nucleosynthesis (Burles et al.~\cite{burles}). The small amount of scatter in observed halo Li abundances is very difficult to reconcile with uniform stellar Li depletions of 0.4\,dex or more. Likewise, it would require improbably high effective temperatures to resolve the disagreement.  Clearly, halo stars and their spectrum formation are not as well understood as perhaps thought.

\section{Discussion}

We have found that while excitation and de-excitation by inelastic H collisions are not important in thermalizing Li, the charge exchange reactions involving the Li($3s$) state, are quite important. This mechanism is almost certainly important also in other alkalis (Na, K, Rb etc), where large cross-sections are also found (Dickinson~et al.~\cite{dickinson}; Janev \&\ Radulovi\'c~\cite{janev}).  For other elements, further detailed calculations will be required to determine whether inelastic H collisions, including excitation and charge exchange, are important.  The main obstacle for such calculations at present is the quantum chemical potentials and couplings required for a large number of molecular states.  Though systems of avoided ionic crossings exist in all metal hydride quasi-molecules, we note other mechanisms may be important.  As suggested by Lambert~(\cite{lambert}), the importance of details in the molecular dynamics, such as avoided ionic crossings, in this process makes a simple general recipe impossible, in contrast to collisional broadening and depolarization by H where such effects play a lesser role  (Anstee \&\ O'Mara~\cite{anstee}; Barklem \&\ O'Mara~\cite{barklem}; Derouich et al.~\cite{derouich}).  {\it Ab initio} potentials for alkaline-earth-hydrides are possible with current techniques (Chambaud \&\ L\'evy~\cite{chambaud_levy}).  However, some cases of particular astrophysical interest such as O+H and Fe+H present a considerable challenge.

\begin{acknowledgements}
We are indebted to F.X. Gad\'ea and A.S. Dickinson for providing us with the quantum chemical data used in computing the cross sections, and their results for mutual neutralization.  We thank D. Kiselman for encouraging us to investigate inelastic H collisions.  We gratefully acknowledge support from the Swedish Research Council, the Swedish Royal Academy of Sciences, the G\"oran Gustafsson Foundation and the Australian Research Council.
\end{acknowledgements}

\end{document}